\documentclass[10pt,pra,aps,notitlepage,superscriptaddress,reprint]{revtex4-1}

\usepackage{listings}

\usepackage[utf8]{inputenc}
\usepackage[T1]{fontenc}
\usepackage[english]{babel}  

\usepackage{tikz}

\usepackage{colortbl}
\usepackage{tabu,diagbox}

\usepackage{pgfplots}
\usepackage{lipsum}
\usepackage{youngtab}
\usepackage{subfigure}

\usepackage{times}
\usepackage{dsfont}

\usepackage{amssymb,amsmath,amsfonts,amsthm}
\usepackage{mathtools}
\usepackage{verbatim}
\usepackage{enumerate}
\usepackage{graphicx}
\graphicspath{figs/}
\usepackage[capitalise]{cleveref}
\usepackage{bbm}
\usepackage{booktabs}

\usepackage{color}
\definecolor{dred}{rgb}{.8,0.2,.2}
\definecolor{ddred}{rgb}{.8,0.5,.5}
\definecolor{dblue}{rgb}{.2,0.2,.8}
\definecolor{dgreen}{rgb}{.2,0.5,.2}
\newcommand{\bra}[1]{\mbox{$\langle #1|$}}
\newcommand{\ket}[1]{\ensuremath{|#1\rangle}}

\newcommand{\tr}{\textrm{tr}}

\DeclareMathOperator{\Tr}{Tr}

\newcommand{\be}{\begin{equation}}

\newcommand{\ee}{\end{equation}}
\newcommand{\bea}{\begin{eqnarray}}
\newcommand{\eea}{\end{eqnarray}}

\usepackage{xcolor}
\usepackage[framed,numbered,final]{matlab-prettifier}

\begin{document}

\title{Determining system Hamiltonian from eigenstate measurements\\ without correlation functions}

\author{Shi-Yao Hou}%
\altaffiliation{These authors contributed equally to this work.}
\affiliation{College of Physics and Electronic Engineering, Center for Computational Sciences,  Sichuan Normal University, Chengdu 610068, China}%
\affiliation{Center for Quantum Computing, Peng Cheng Laboratory, Shenzhen, 518055, China}

\author{Ningping Cao}
\altaffiliation{These authors contributed equally to this work.}
\affiliation{Department of Mathematics \& Statistics, University of Guelph, Guelph N1G 2W1, Ontario, Canada}%
\affiliation{Institute for Quantum Computing, University of Waterloo, Waterloo N2L 3G1, Ontario, Canada}

\author{Sirui Lu}%
\affiliation{Department of Physics, Tsinghua University, Beijing 100084, China}%

\author{Yi Shen}%
\affiliation{Department of Statistics and Actuarial Science,
  University of Waterloo, Waterloo, Ontario, Canada}%

\author{Yiu-Tung Poon}%
\affiliation{Department of Mathematics, Iowa State University, Ames, Iowa, IA 50011, USA.}%
\affiliation{Center for Quantum Computing, Peng Cheng Laboratory, Shenzhen, 518055, China}

\author{Bei Zeng}
\email{zengb@ust.hk}
\affiliation{Department of Physics, The Hong Kong University of Science and Technology, Clear Water Bay, Kowloon, Hong Kong, China}
\affiliation{Department of Mathematics \& Statistics, University of Guelph, Guelph N1G 2W1, Ontario, Canada}%
\affiliation{Institute for Quantum Computing, University of Waterloo, Waterloo N2L 3G1, Ontario, Canada}

\date{\today}

\begin{abstract}
Local Hamiltonians arise naturally in physical systems. Despite its seemingly `simple' local structure, exotic features such as nonlocal correlations and topological orders exhibit in eigenstates of these systems. 
Previous studies for recovering local Hamiltonians from measurements on an eigenstate $\ket{\psi}$ require information of nonlocal correlation functions. In this work, we argue that local measurements on $\ket{\psi}$ is enough to recover the Hamiltonian in most of the cases.
Specially, we develop an algorithm to demonstrate the observation.
Our algorithm is tested numerically for randomly generated local Hamiltonians of different system sizes and returns promising reconstructions with desired accuracy. Additionally, for random generated Hamiltonians (not necessarily local), our algorithm also provides precise estimations. 

\end{abstract}

\maketitle

\section{Introduction}

The principle of locality, arising naturally in physical systems, states that objects are only affected by their nearby surroundings.
Locality is naturally embedded in numerous physical systems characterized by local Hamiltonians, which play a critical role in various quantum physics topics, such as quantum lattice models~\cite{tarasov2016local,tarasov1985irreducible,zanardi2002quantum}, quantum simulation~\cite{feynman1982simulating,cirac2012goals,lloyd1996universal,buluta2009quantum}, topological quantum computation~\cite{freedman2003topological}, adiabatic quantum computation~\cite{nagaj2008local,aharonov2008adiabatic,jordan2006error},  and quantum Hamiltonian complexity ~\cite{kempe2006complexity,kempe20033,bravyi2006complexity}.
In the past few years, rapidly developing machine learning techniques allow us to study these topics in a new manner~\cite{xin2018local,bairey2019learning,deng2017machine,PhysRevB.99.155136,pudenz2013quantum,liu2017self}.
Empowered by traditional optimization methods and contemporary machine learning techniques, we map the task of revealing the information encoded in a single eigenstate of a local Hamiltonian to an optimization problem.

For a local Hamiltonian $H=\sum_i c_iA_i$ with $A_i$ being some local operators, it is known that a single (non-degenerate) eigenstate $\ket{\psi}$ can encode the full Hamiltonian $H$ in certain cases~\cite{garrison2018does,qi2017determining,chen2012correlations}, 
 such as when the expectation value of $A_i$s on $\ket{\psi}$,  $a_i=\bra{\psi}A_i\ket{\psi}$, are given and further assumptions are satisfied. 
A simple case is that when $\ket{\psi}$ is the unique ground state of $H$; thus the corresponding density matrix of $\ket{\psi}$ can be represented in the thermal form as
\begin{equation}
\label{eq:H}
\ket{\psi}\bra{\psi}=\frac{e^{-\beta H}}{\tr(e^{-\beta H})}
\end{equation}
for sufficiently large $\beta$. This implies that the Hamiltonian $H$ can be directly related to $\ket{\psi}$, and hence can be determined by the measurement results $a_i$, using algorithms developed in the literature ~\cite{zhou2008irreducible,niekamp2013computing} for practical cases.
Because $A_i$s are local operators, the number of parameters of $H$ (i.e., the number of $c_i$s) is only polynomial in terms of system size. We remark that the problem of finding $H$ 
is also closely related to the problem of determining quantum states from local measurements~\cite{xin2018local,bairey2019learning,kalev2015power, linden2002almost, linden2002parts, diosi2004three, chen2012comment, chen2012ground, chen2013uniqueness,zeng2015quantum}, and also has a natural connection to the study of quantum marginal problem~\cite{Kly06,Liu06,LCV07,WMN10}, as well as its bosonic/fermionic version that are called the $N$-representability problem~\cite{Col63,Erd72,Kly04,Kly06,AK08,SGC13,walter2013entanglement,sawicki2014convexity}. 

For a wavefunction $\ket{\psi}$ that is an eigenstate (i.e. not necessarily a ground state), one interesting situation is related to the eigenstate thermalization hypothesis (ETH)~\cite{Deutsch1991eth,Srednicki1994eth,Srednicki1996eth,Srednicki1999eth,luca2016eth}. When the ETH is satisfied, the reduced density matrix of a pure and finite energy density eigenstate for a small subsystem becomes equal to a thermal reduced density matrix asymptotically~\cite{garrison2018does}. 
In other words, Eq.~\eqref{eq:H} will hold for some eigenstate $\ket{\psi}$ of the system in this case, and one can use a similar algorithm~\cite{zhou2008irreducible,niekamp2013computing} to find $H$ from $a_i$s, as in the case of ground states. Another situation previously discussed is that if the two-point correlation functions $\bra{\psi}A_iA_j\ket{\psi}$ are known, one can reproduce $H$ without satisfying ETH~\cite{chen2012correlations,qi2017determining}. 
Once the two-point correlation functions are known, one can again use an algorithm to recover $H$ from the correlation functions, for the case of ground states.
However, in practice, the nonlocal correlation functions $\bra{\psi}A_iA_j\ket{\psi}$ are not easy to obtain~\cite{li2017optimal}. 

In this paper, we answer a simple but significant question: can we determine a local Hamiltonian ($c_i$s) from only the local information ($a_i$s) of any one of the eigenstates ($\ket{\psi}$) without further assumptions. 
Obviously, there are cases that an eigenstate can not determine the system Hamiltonian, such as the product state and the eigenstates of a frustration-free Hamiltonians.
However, for a randomly chosen physical system of which the Hamiltonian has no special structures, 
we show that only the knowledge of $a_i$s is sufficient to determine $H$.

\begin{figure*}[t]
\includegraphics[width=.7\linewidth]{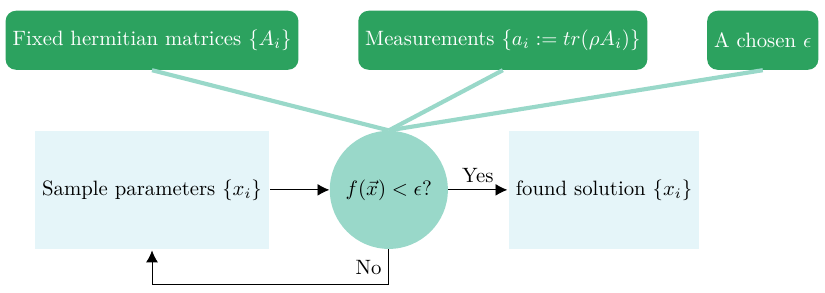}
\caption{\textbf{Diagram of our algorithm: }The rounded boxes (the first row) represent known information - the fixed set of Hermitian operators $\{A_i\}$, measurement outcome $\{a_i\}$ and a chosen precision parameter $\epsilon$. The information in hand contributes to the objective function $f$~(\cref{eq:obj}). The second row demonstrates the procedure of the algorithm: first, sample a set of random parameters $\vec x = \{x_i\}$; second, estimate the objective function $f(\vec x)$ and determine whether $f(\vec x)$ is smaller then the chosen $\epsilon$; update $\vec x$ until it satisfies $f(\vec x) < \epsilon$.}\label{fig:diagram}
\end{figure*}

Based on the knowledge in hand - a set of possible Hermitian operators $\{A_i\}$ for the system Hamiltonian and the measurement outcomes $\{a_i\}$, we formulate a positive-semidefinite function $f(\vec x)$~(\cref{eq:obj}), with the $f(\vec{c})=0$ for the desired $\vec{c}$. 
The problem of finding the exact Hamiltonian converts to a unconstrained optimization problem of finding $\vec x$ that minimizing $f$. 
A small real number $\epsilon$ is chosen to control the precision of the result.
We update the sampled $\vec x$ until $f(\vec x)$ is less than $\epsilon$.
\cref{fig:diagram} depicts the procedure of our algorithm. 

We test the algorithm in two scenarios: one with four randomly generated operator $A_i$s acting on the whole system, and one with two random local operators $A_i$s. 
Our algorithm almost perfectly reproduces $c_i$s - the average fidelities for both cases are close to $1$.
Since the algorithm recovers Hamiltonians with almost perfect fidelities
Based on the reconstructed Hamiltonian $H$ (i.e. $c_i$s), one can also recover the eigenvalue of $\ket{\psi}$ from $c_i$s and $a_i$s, and the wave function itself from the eigenvectors of Hamiltonian $H$.
In the case when $A_i$s are local operators, our method can find $c_i$s from only local measurement results, hence shed light on the correlation structures of eigenstates of local Hamiltonians.

\section{Algorithm}
\label{sec:methods}

We start to discuss our method in a general situation, where the Hamiltonian $H$ can be expressed in terms of a set of known Hermitian operators $\{A_1,A_2,\ldots,A_m\}$: $H=\sum c_i A_i$ with $(c_1,c_2,\ldots,c_m)=\vec{c}$. 
For an eigenstate $\ket{\psi}$ of $H$ with unknown eigenvalue $\lambda$, which satisfies $H|\psi\rangle=\lambda |\psi\rangle$, we can denote the measurement results as $a_i=\langle\psi|A_i|\psi\rangle$. 
With only knowing the measurement results $a_i$s, our goal is to find the coefficients $\vec{c}$ to determine $H$.

We observe that, even if $|\psi\rangle$ is not the ground state of $H$, it can be the ground state of another Hamiltonian  $\tilde{H}^2$ with $\tilde{H}$ given by
\begin{equation}
\tilde{H}=H-\lambda I=\sum_i c_i(A_i-a_i I),
\end{equation}
since 
\begin{equation}
\bra{\psi} H\ket{\psi}=\lambda=\sum_i c_i \bra{\psi}A_i\ket{\psi}=\sum_i c_ia_i.
\end{equation}

Then the density matrix of $\ket{\psi}$ (which is in fact of rank 1) can be written in the form of a thermal state:
\begin{equation}
\rho(\vec{c})=\ket{\psi}\bra{\psi}=\frac{e^{-\beta\tilde{H}^2}}{\tr(e^{-\beta\tilde{H}^2})}
\end{equation}
for sufficiently large $\beta$, satisfying
\begin{equation}
\tr(A_i\rho(\vec{c}))=a_i,\ \tr(\tilde{H}^2\rho(\vec{c}))=0.
\end{equation}

With these conditions in mind, we are ready to reformulate our task as an optimization problem with the following objective function: 

\begin{equation}\label{eq:obj}
f(\vec{x})=\sum_{i=1}^{m}[\tr(A_i\rho(\vec{x}))-a_i]^2+ \tr(\tilde{H}^2\rho(\vec{x})),
\end{equation}
where $\vec{x}$ is the estimation of $\vec{c}$, and $\vec{x} = \vec{c}$ when $f(\vec{x})$ is minimized.

Notice that the first term of $f(\vec{x})$ is minimized by $\tr(A_i\rho(\vec{c}))=a_i$, which guarantees that the state $\rho(\vec{c})$ obtained is the state that produces the desired measurement outcomes on $A_i$s. However, there may be many (thermal) states which also yield such outcomes. By simply minimizing this first term, optimization algorithms tend to return thermal states $\rho(\vec{c})$ with nonzero entropy, which is not the eigenstate of $H$ (with entropy zero) that we are willing to find. In order to fix this issue and ensure that the optimization returning a (near) rank $1$ state $\rho(\vec{c})$, we add the second term, which is only zero unless $\rho(\vec{c})$ is the ground state of $\tilde{H}^2$, hence an eigenstate of $H$.
Combining these two terms together, we make sure that when the minimum value of  Hamiltonian  $f(\vec{x})$ is reached, we will obtain a $\rho(\vec{c})$ corresponding to measurement outcomes $a_i$, and at the same time an eigenstate of $H$. In practice, we set up a parameter $\epsilon$, such that with $f(\vec{x})<\epsilon$, we find a result with high fidelity to the desired value of $\{c_i\}$.

For the convenience of numerical implementation, we let $\tilde{H}_\beta = \sqrt{\beta}\tilde{H}$, then the thermal state $\rho(\vec{c})$ becomes
\begin{equation}
\rho(\vec{c})=\frac{e^{-\tilde{H}_\beta^2}}{\tr(e^{-\tilde{H}_\beta^2})}.
\end{equation}
Consequently, the objective function~\cref{eq:obj} can be rewritten in an equivalent form
\begin{equation}
f(\vec{x})=\sum_{i=1}^{m}[\tr(A_i\cdot\rho(\vec{x})]-a_i)^2+ \tr(\tilde{H}_\beta^2\cdot\rho(\vec{x})).
\label{eq:fc}
\end{equation}
We aim to solve for $f(\vec{x})=0$ by minimizing $f(\vec{x})$. In practice, we terminate our iterations when $f(\vec{c})$ is smaller than a fixed small value $\epsilon$. The corresponding optimization result is denoted as $\vec c_\text{opt}$.
As we reformed the objective function by using $\tilde{H}_\beta$, the result is actually $\vec c_\text{opt}=\sqrt\beta\cdot\vec c$.

Theoretically, we need that $\beta$ turns to infinity to pick up the ground state of $\tilde{H}_\beta^2$. In practice, however, we only require that $\beta$ is some ``large number". What is more, to pick up the ground state of $\tilde{H}_\beta^2$ for $\rho(\vec{c})$, what really matters is in fact the gap between the first excited state and the ground state of $\tilde{H}_\beta^2$. 
Therefore, we just simply set $\beta=1$ and let the optimizer automatically amplify the energy gap during iteration, when $\rho(\vec{c})$ is approaching the desired state.
A more detailed discussion regarding the choice of $\beta$ can be found in ~\cref{sec:app}. 

Since all the constraints are written in~\cref{eq:fc}, minimizing $f(\vec{c})$ is an unconstrained minimization problem. There are plenty of standard algorithms for this task. In our setting, computing the second-order derivative information of $f(\vec{c})$  is quite complicated and expensive. 
Therefore, instead of using Newton method which requires the Hessian matrix of the second derivatives of the objective function, we choose the quasi-Newton method with BFGS-formula to approximate the Hessians~\cite{bfgs_b,bfgs_f,bfgs_g,bfgs_s}.
The MATLAB function~\textsc{fminunc}, which uses the quasi-Newton algorithm, is capable of realizing this algorithm starting from an initial random guess of $c_i$s. 
When the quasi-Newton algorithm fails (converges into a local minimum), we start with a different set of random initial value $c_i$s and optimize again until we obtain a good enough solution. 

The BFGS algorithm is a typical optimization algorithm which requires gradients of the objective function on $c_i$s. The form of the objective function $f(\vec{c})$ is so complicated such that computing the gradient is a difficult task. 
To solve this issue, we borrow the methods of computational graph and auto differentiation from machine learning. The computational graph is shown in~\cref{fig:comgra}, in which we show the intermediate functions and variables. Mathematically, the final gradients could be calculated via chain rules. However, since some of the intermediate variables are matrices and complex-valued, the automatic differentiation toolboxes, which deal with real variables, can not be applied directly. To obtain the gradients, we have to careful handle the derivation of the intermediate functions, especially those with matrices as their variables. More details about the our gradient method can be found in Appendix~\ref{sec:grad}. 

We use Matlab to implement our algorithm, and the detailed implementation can be found in Appendix~C.

\section{Results}

In this section, we test our algorithm in three steps as follows. 
First, we randomly generate several $A_i$s and $c_i$s, hence the Hamiltonian $H_{rd}$ and its eigenstates. 
Second, for each Hamiltonian, we randomly choose one eigenstate $|\psi\rangle$, therefore we have $a_i=\langle \psi|A_i|\psi\rangle$ and $\rho=|\psi\rangle\langle\psi|$. 
Hereby we can run our algorithm to find the Hamiltonian $H_{al}$. 
Comparing $H_{al}$ and $H_{rd}$, we then know that how well the approach works. 
The algorithm has been tested for two cases: $A_i$ being the generic operator and local operator.

To compare $H_{al}$ and $H_{rd}$, we need a measure to characterize the similarity, or \textit{distance} between these two Hamiltonians. 
The metric we used here is the following fidelity as discussed in~\cite{fortunato2002design}: 
\begin{equation}
    f(H_{al},H_{rd})=\frac{\tr H_{al} H_{rd}}{\sqrt{\Tr H_{al}^2} \sqrt{\Tr H_{rd}^2}}.
    \label{eq:fidelity}
\end{equation}
To see the meaning of this metric, 
notice that $\tr H_{al} H_{rd}$ is the inner product of the two Hamiltonians, while $\sqrt{\Tr H_{al}^2}$ and $\sqrt{\Tr H_{rd}^2}$ are the two normalization constant. Therefore, $f(H_{al},H_{rd})\in [0,1]$. If $H_{al}$ and $H_{rd}$ describe the same system up to a constant $b\in \mathbb{R}$, then $f(H_{al},H_{rd})=1$. Smaller $f(H_{al},H_{rd})$ then indicates that the two Hamiltonians are more far apart. Moreover, notice that in our settings the Hamiltonians are represented by vectors $\vec{c}$ and $\vec{c}'$ in $m$-dimensional real space. If the chosen $A_i$s are normalized and orthogonal, which means
\begin{equation}
\tr(A_iA_j)=d\delta_{ij},
\end{equation}
where $d$ is the normalization constant given by the system dimension (e.g. for Pauli matrices, $d=2$), this fidelity definition is exactly the cosine loss function of $\vec{c}$ and $\vec{c}'$, where $\vec{c}'$ is generated from our algorithm, that is, for normalized $A_i$s,
\begin{equation}
    f(H_{al},H_{rd})=\frac{\Tr H_{al} H_{rd}}{\sqrt{\Tr H_{al}^2} \sqrt{\Tr H_{rd}^2}}=\frac{\vec{c}\cdot\vec{c}'}{\parallel \vec{c}\parallel\parallel \vec{c}'\parallel},
    \label{eq:fidelity_1}
\end{equation}
where $\parallel \vec{c}\parallel$ means the 2-norm of the vector $\vec{c}$.

\subsection{Results with general operators}

When implementing our method on generic operators, we randomly generate three Hermitian operators $A_i$s and fix them. 
We also create several real constants $c_i$s randomly, then assemble $A_i$s and $c_i$s into Hamiltonians
\begin{equation}
H=c_1A_1+c_2A_2+c_3A_3.
\end{equation}
After diagonalizing $H$, we choose one eigenvector $|\psi\rangle$ and calculate the expectation values $a_i=\langle\psi|A_i|\psi\rangle$.

\begin{figure}[t]
\subfigure[4 qubits]{\includegraphics[scale=0.25]{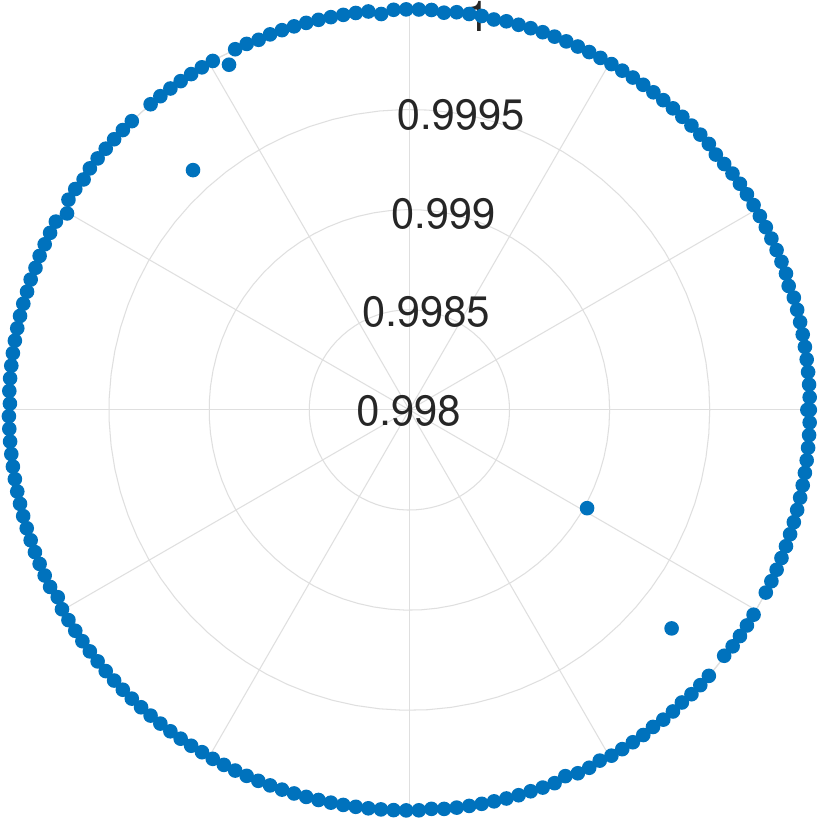}} \qquad      
\subfigure[5 qubits]{\includegraphics[scale=0.25]{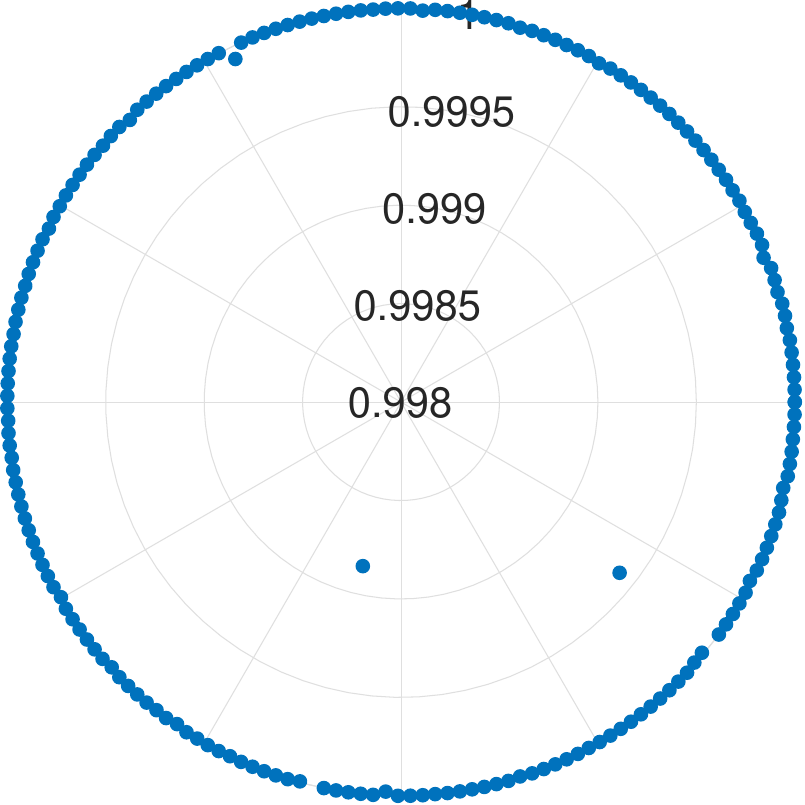}} \qquad
\subfigure[6 qubits]{\includegraphics[scale=0.25]{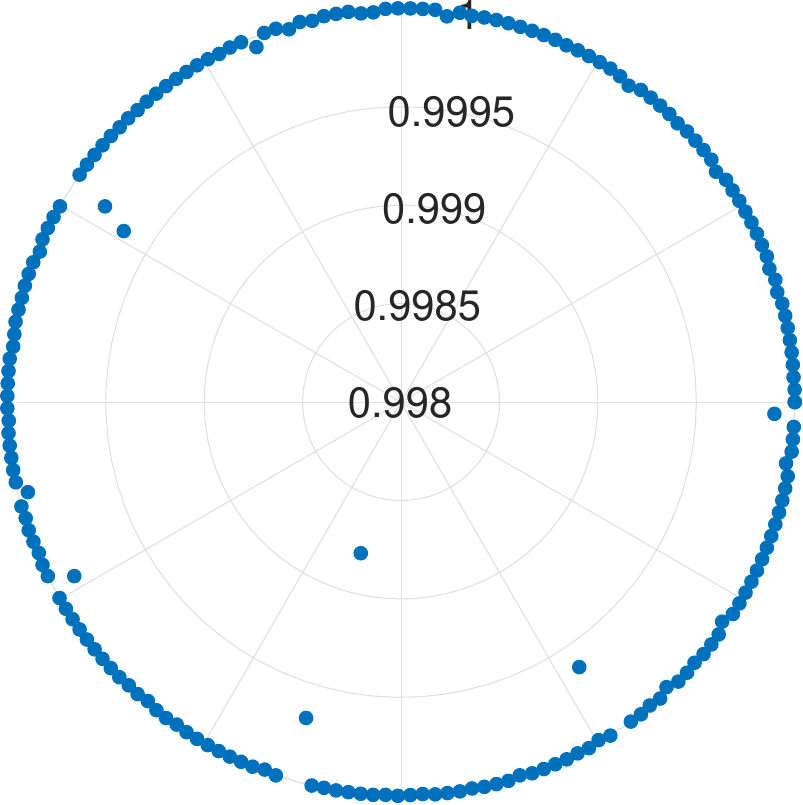}} \qquad         
\subfigure[7 qubits]{\includegraphics[scale=0.25]{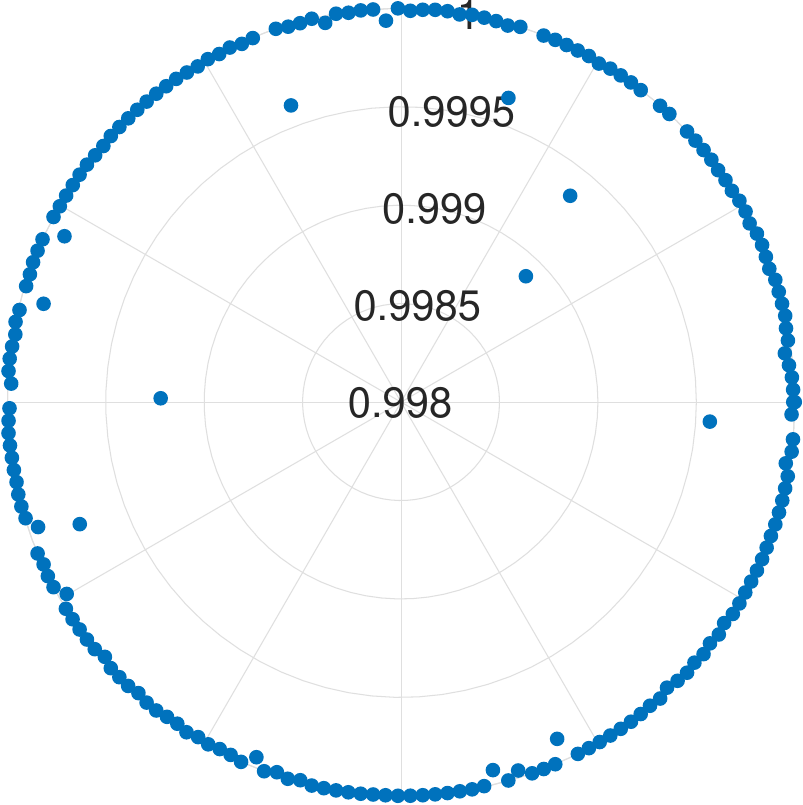}}

\caption{\textbf{Results for generic operators.} Each dot represents the fidelity $f$ of a test data point. (a) 4 qubits. (b) 5 qubits. (c) 6 qubits. (d) 7 qubits. }
\label{fig:gen_res}
\end{figure}

For $n$-qubit systems, the dimension of the system is $d=2^n$.
We tested the cases for $n=4, 5, 6$ and $7$. 
For each $n$, we generate $200$ data points. 
Each test set is $\{c_i,A_i, a_i|i=1,2,3\}$, where $c_i\in (0,1)$ for $i=1,2,3$ and $A_i$s are randomly generated Hermitian matrices of dimension $d=2^n$ 
\footnote{To generate a random Hermitian matrix, we first generate $d^2$ complex number $\{e_{ij}|0\leq i,j\leq d\}$ as the entries of a matrix $E$, then we construct $A=E+E^{\dagger}$. It can be easily seen $A$ is an Hermitian matrix.}. 
With these data, $H_{rd}=\sum_i c_i A_i$ is obtained. 
Diagonalizing $H_{rd}$ and randomly choosing one eigenvector $|\psi\rangle$, we obtain $\{a_1,a_2,a_3\}$. We show the results for applying our algorithm to all cases in~\cref{fig:gen_res}.

We find that the final fidelities are larger than 99.8\% for all tested cases. Although the fidelity slightly decreases with the increasing number of qubits, the lowest fidelity (7-qubit case) is still higher than 99.8\%. 
Because the eigenvector $|\psi\rangle$ is randomly chosen, our method is not dependent on the energy level of eigenstates.

\subsection{Results with local operators}

\begin{figure*}[!htb]
\subfigure[]{
\begin{minipage}{0.25\linewidth}
\includegraphics[width=0.5\linewidth]{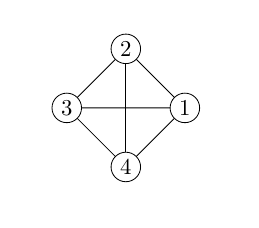}
\end{minipage}
}
\subfigure[]{
\begin{minipage}{0.3\linewidth}
\includegraphics[width=0.7\linewidth]{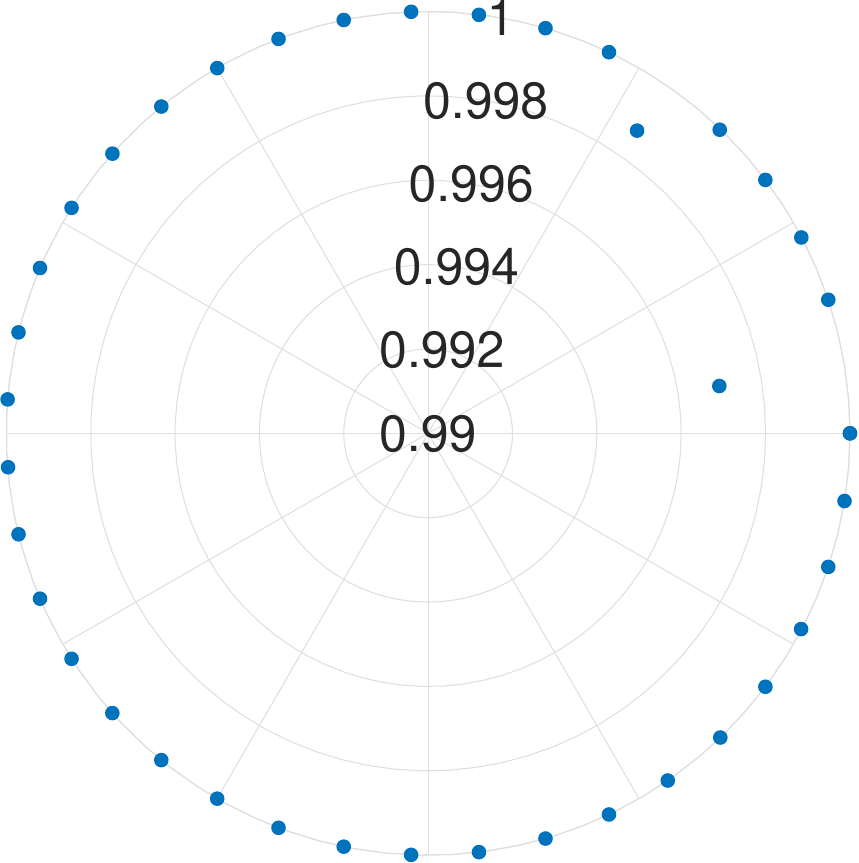}
\end{minipage}
}
\subfigure[]{
\begin{minipage}{0.4\linewidth}
\includegraphics[width=1\linewidth]{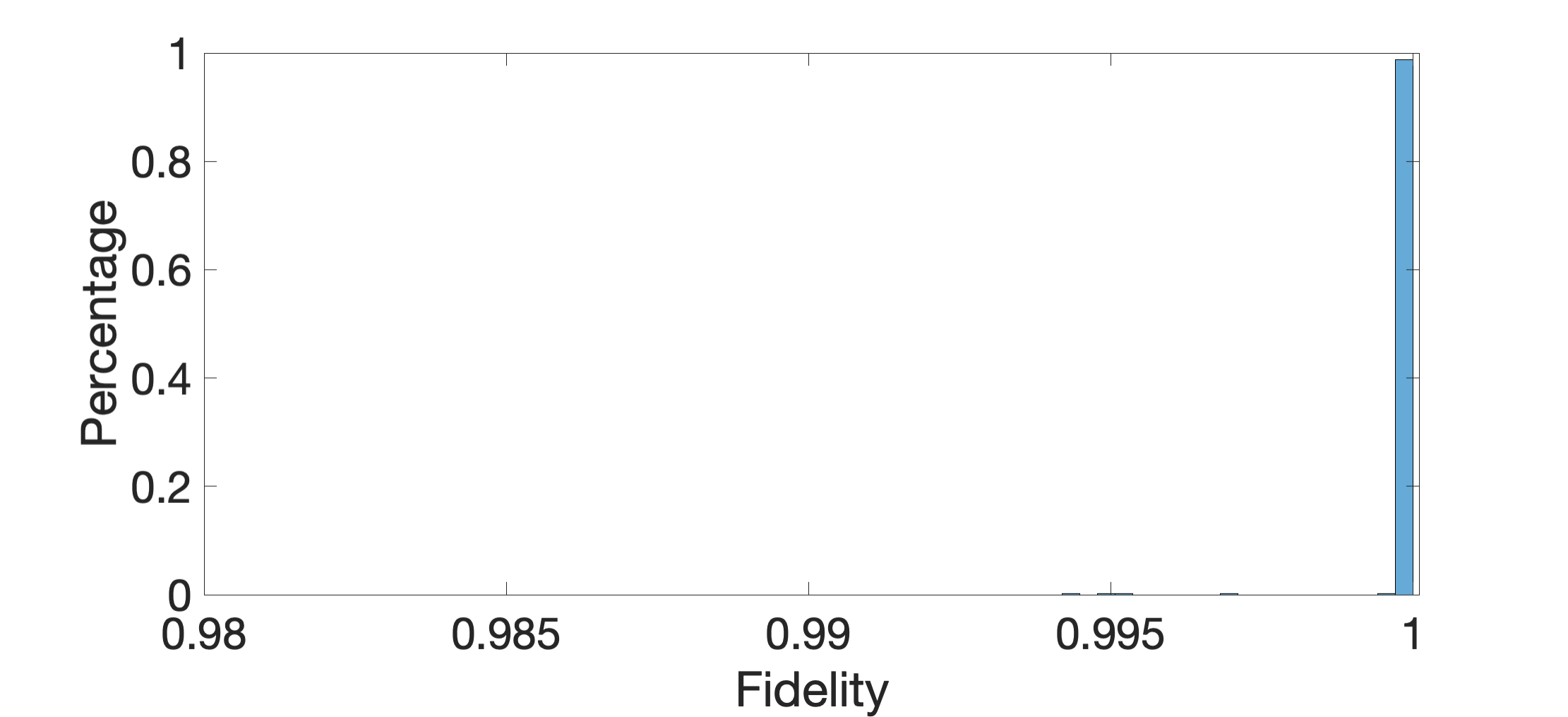}
\end{minipage}
}

\subfigure[]{
\begin{minipage}{0.25\linewidth}
\includegraphics[width=1\linewidth]{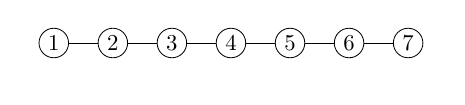}
\end{minipage}
}
\subfigure[]{
\begin{minipage}{0.3\linewidth}
\includegraphics[width=0.7\linewidth]{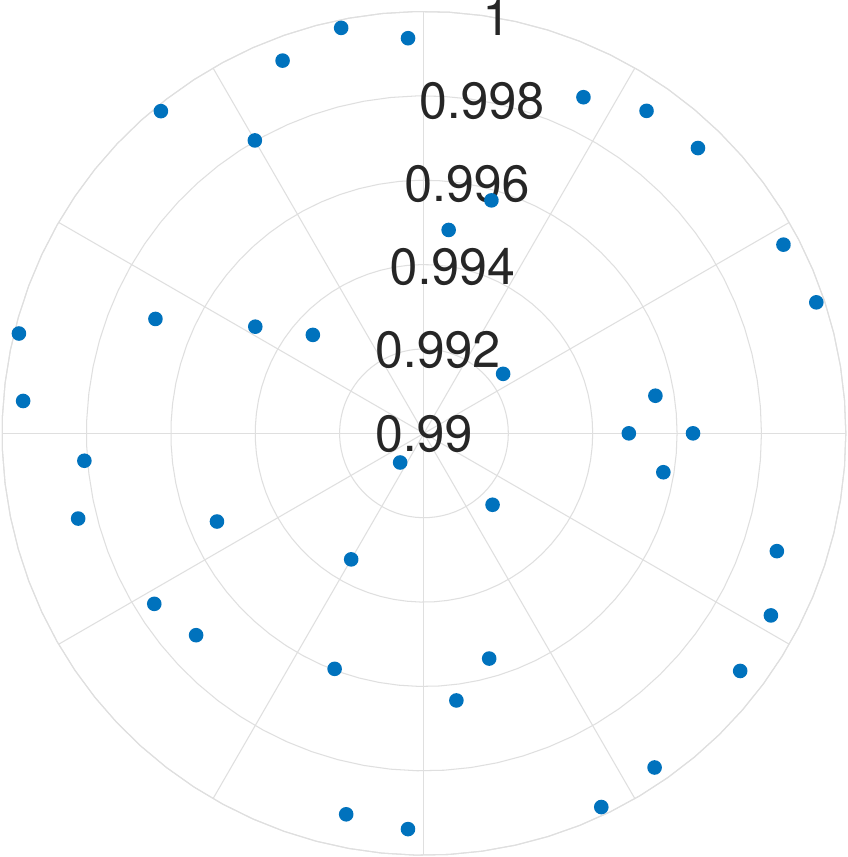}
\end{minipage}
}
\subfigure[]{
\begin{minipage}{0.4\linewidth}
\includegraphics[width=1\linewidth]{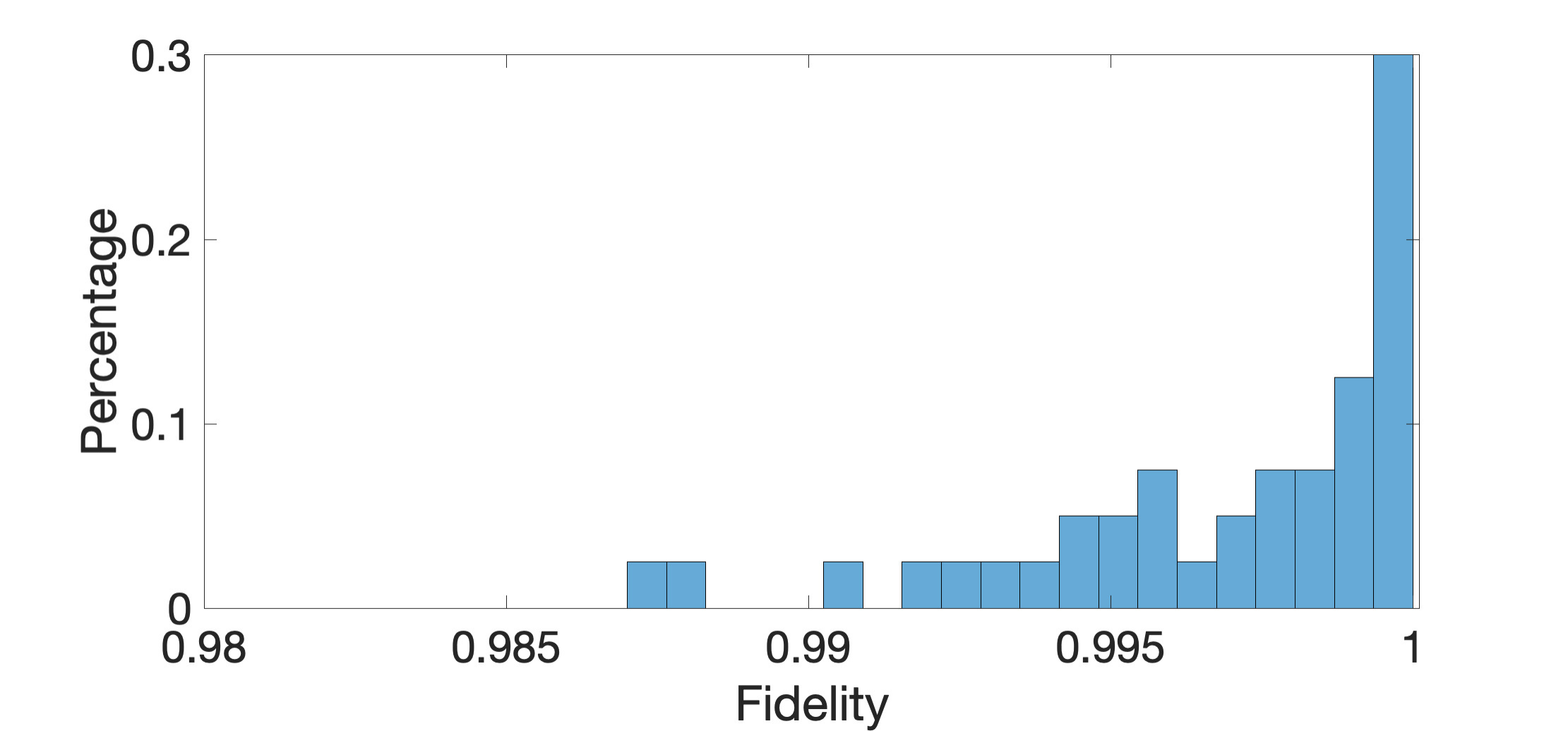}
\end{minipage}
}

\caption{\textbf{Configurations and Results of tested local Hamiltonians} (a) Structure of a four-qubit full connected lattice. (b) The fidelity $f$ between random  $4$-qubit fully connected lattice Hamiltonians $H_{rd}$ and Hamiltonian obtained in our algorithm $H_{al}$. (c) The fidelity distribution of $4$-qubit systems.  (d) Structure of a $7$-qubit chain lattice (e) The fidelity $f$ between $7$-qubit $2$-local Hamiltonians $H_{rd}$ and Hamiltonian estimation $H_{al}$ of our algorithm. (f) The fidelity distribution of our $7$-qubit systems.}
\label{fig:res_local}
\end{figure*}

In this section, we report our results on the systems with a $2$-local interaction structure. 
We tested two different structures, shown in~\cref{fig:res_local}(a) and~\cref{fig:res_local}(d). Each circle represents a qubit on a lattice, and each line represents an interaction between the connected two qubits.

The fully-connected $4$-qubit system is shown in~\cref{fig:res_local}(a). 
The Hamiltonian can be written as 
\begin{equation}
H=\sum_{1\le j,i \leq 4}\sum_{i < j} c_{ij} A_{ij},
\end{equation} 
where $c_{ij}$s are real parameters and $A_{ij}$s are random generated $2$-local operators. One eigenstate out of 16 is randomly chosen as the state $|\psi\rangle$. 
Our algorithm has been tested on 800 such Hamiltonians. 

We then analyze the $7$-qubit chain model shown in~\cref{fig:res_local}(d). Similarly, we can write the Hamiltonian as 
\begin{equation}
H=\sum_{1\le i\leq 6}c_{i,i+1} A_{i,i+1},
\end{equation}
where $c_{i,i+1}$s and $A_{i,i+1}$s are the parameters and 2-local interactions. 
We randomly generate $40$ such Hamiltonians and applied our algorithm.

The results of these two $2$-local Hamiltonians are shown in~\cref{fig:res_local}. Our algorithm recovered Hamiltonians with high fidelities for both cases. 
The average fidelity for our 4-qubit (7-qubit) system is 99.99\% (99.73\%).
As the dimension of the system increases, the fidelity between $H_{rd}$ and $H_{al}$ is slightly decreased.
The histogram of the fidelities shows that, for most data points, the fidelities are very close to $1$. 
Our algorithm almost perfectly recovered these $2$-local Hamiltonian from measurement outcomes of a randomly picked eigenvector.

\begin{figure}[ht]
\includegraphics[width =  \linewidth]{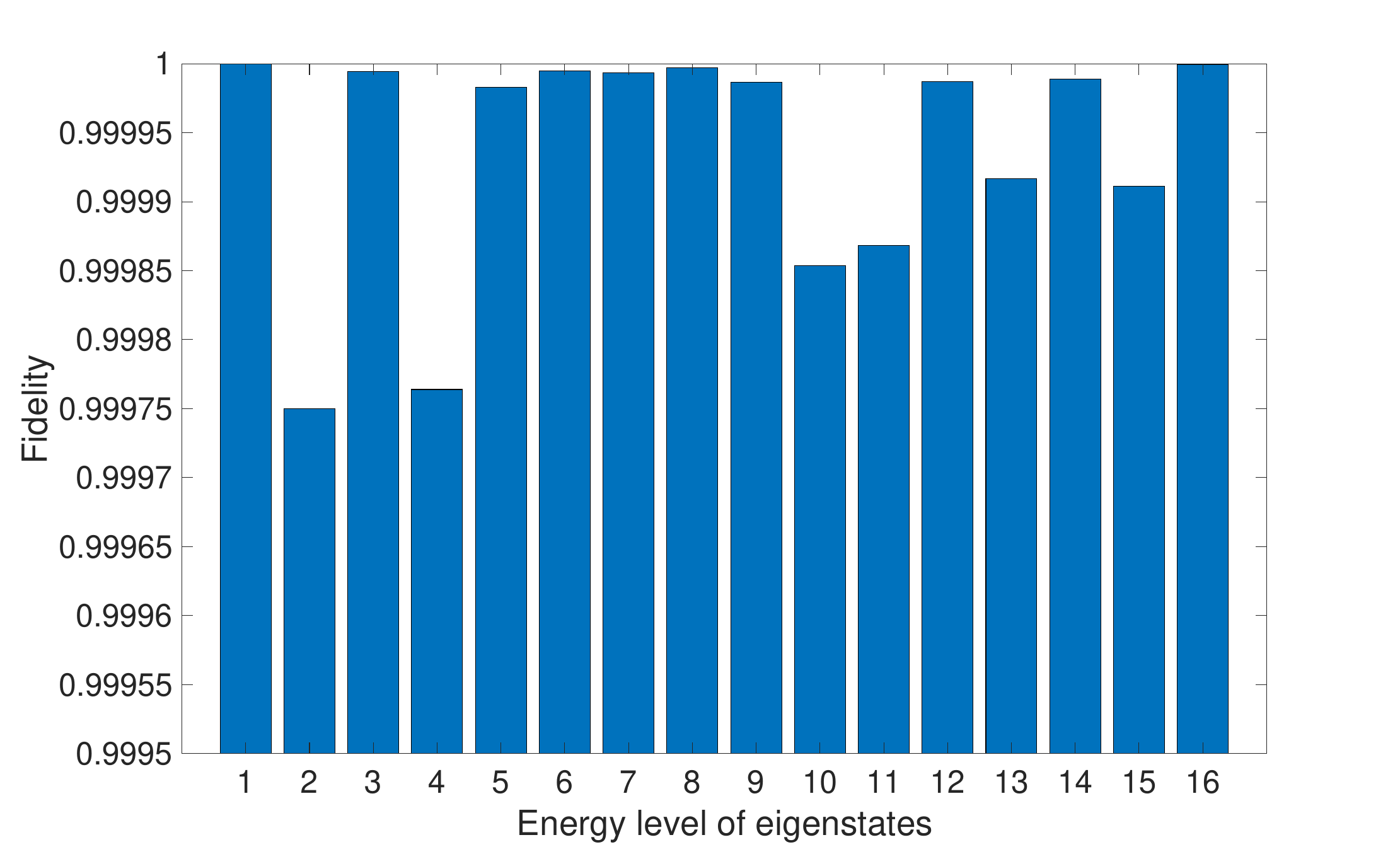}
\caption{\textbf{The average fidelities $f$ by applying our algorithm to different energy levels of eigenstates of the $4$-qubit case.} 
The average fidelities for different energy levels of eigenstates are all higher than 99.97\%.}
\label{fig:res_4_ind}
\end{figure}

We examine the effectiveness of our algorithm according to different eigenvectors of the same Hamiltonians. 
\cref{fig:res_4_ind} demonstrates average fidelities between $H_{rd}$ and $H_{al}$ of $4$-qubit case by chosen different eigenvectors. 
These average fidelities are higher than 99.9\% for all $16$ eigenvectors.
Hence, the effectiveness of our algorithm is independent of energy levels. 

\section{Further analysis of the algorithm}

In this section, we analyze the error tolerance and the performance of the algorithm, based on the results of our numerical experiments.

\subsection{Error tolerance analysis}

The numerical tests in previous sections deal with noiseless theoretical data. In practical scenarios, however, data is always noisy. Here we provide analysis of error tolerance for our algorithm.

As an example, we consider a 4-qubit system with Hamiltonian $H=\sum_{i=1}^{3}c_i A_i$
where $A_i$'s are random generated $16$ by $16$ Hermitian operators. Choose one eigenstate $\ket{\psi}$ of $H$, the noiseless measurements of the eigenstate are denoted as $\{a_i | a_i=\bra{\psi}A_i\ket{\psi}\}$. Noises used here are randomly drawn from normal distributions $\gamma \sim \mathcal{N}(0, \sigma^2)$. Adding the generated noise $\gamma$ to measurements $a_i$, the noisy data $a_{i,m}$ follows the normal distribution $\mathcal{N}(a_i, \sigma^2)$. 

We can control how noisy $a_{i,m}$ is by changing the standard deviation $\sigma$. 
Clearly the noisiness of data is relative to the magnitude of true values, thus $\sigma/a_{i}$ can be used as an indicator.
We test ten different $\sigma$'s of which $\sigma/a_{i}$'s ranging from $0.01$ (1\%) to $0.1$ (10\%). For each $\sigma$, $1000$ data points have been generated. 

\begin{figure}[ht]
\includegraphics[width = \linewidth]{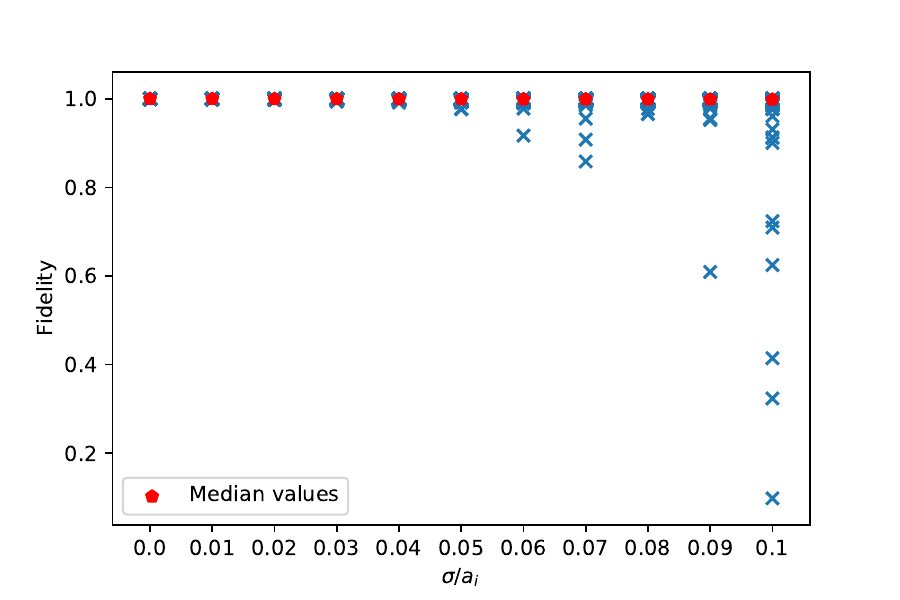}
\caption{\textbf{Fidelities of noisy data:} Each $\sigma/a_i$ ($\sigma$) has 1000 test points. Blue cross marks represent the fidelities between Hamiltonians recovered with noisy data $H_\text{err}$ and real Hamiltonians $H$ . The red pentagons are median values of fidelities for each $\sigma/a_i$. }
\label{fig:error}
\end{figure}

The Hamiltonians attained by our algorithm with these noisy data $a_{i,m}$ are denoted as $H_\text{err}$. The fidelity between each pair of the true Hamiltonian $H$ and $H_\text{err}$ is shown in~\cref{fig:error}. Though increasing the noise cause the fidelities of a few points significantly decreased, all median fidelities of different $\sigma$ are above 99.8\%. Even for $\sigma/a_{i} = 0.1$ (added 10\% of noise), only 2.1\% of data points have fidelities lower than 99.0\%. In other words, singular points which have significant low fidelities are rare. This is demonstrates that the performance of our algorithm is stable.

\subsection{Performance analysis}

The convergence and time cost of our method are analyzed in this subsection.
We study the relation between the halting condition $\epsilon$ and the convergence of our algorithm. 
We also observed that the time cost tends to differ for each Hamiltonian configuration.

\begin{figure*}[t]
\begin{minipage}{0.47\textwidth}
\centering
\includegraphics[width = \linewidth]{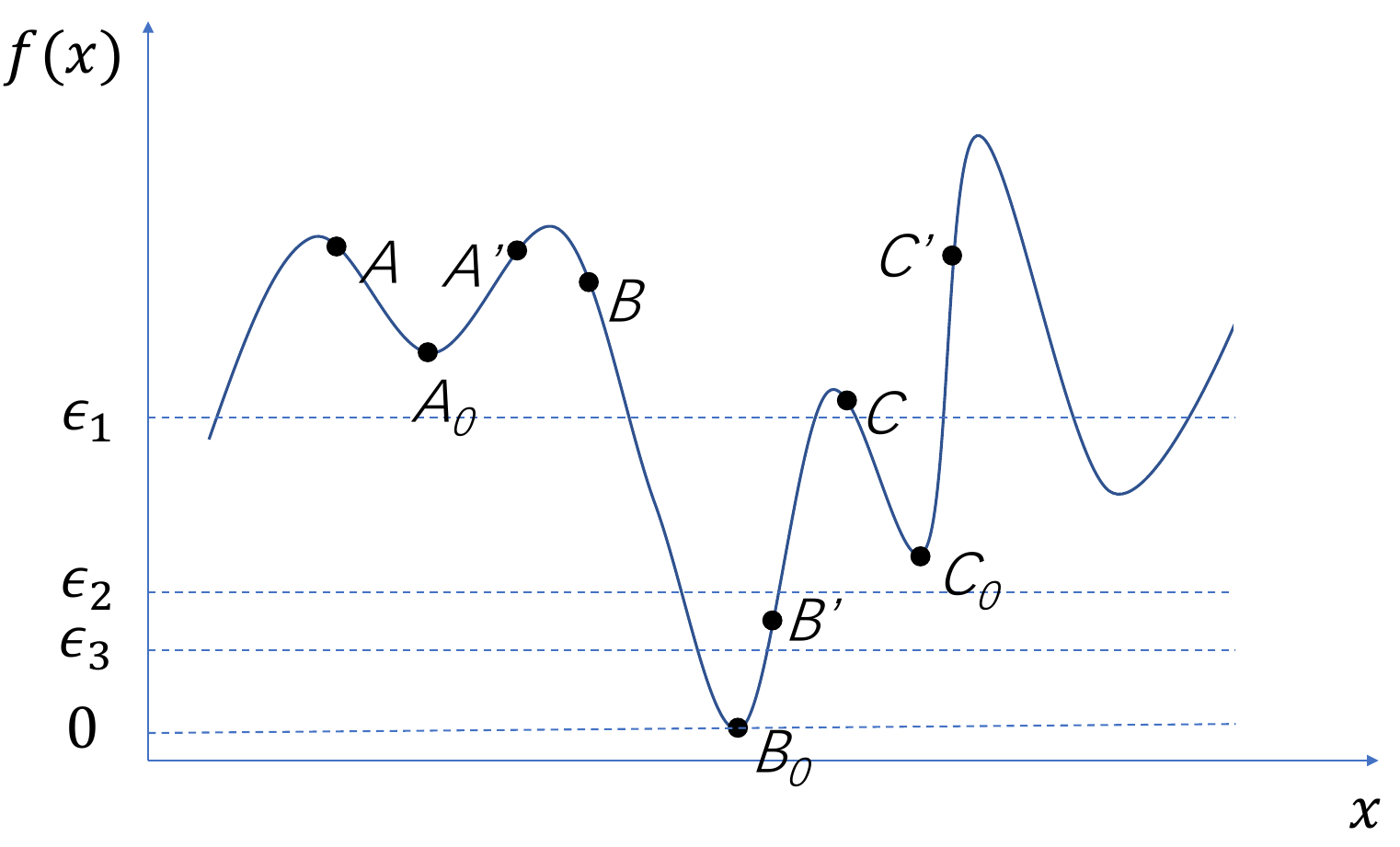}
\caption{\textbf{The schematic diagram for demonstrate the influences of local minima.} $A_0$ and $C_0$ are local minima of the object function $f(x)$. $B_0$ is the global minimum. A slightly larger $\epsilon$ may cause the iteration sticks in the local minima.}
\label{fig:mindemo}
\end{minipage}
$\quad$
\begin{minipage}{0.47\textwidth}
\centering
\includegraphics[width = 0.85\linewidth]{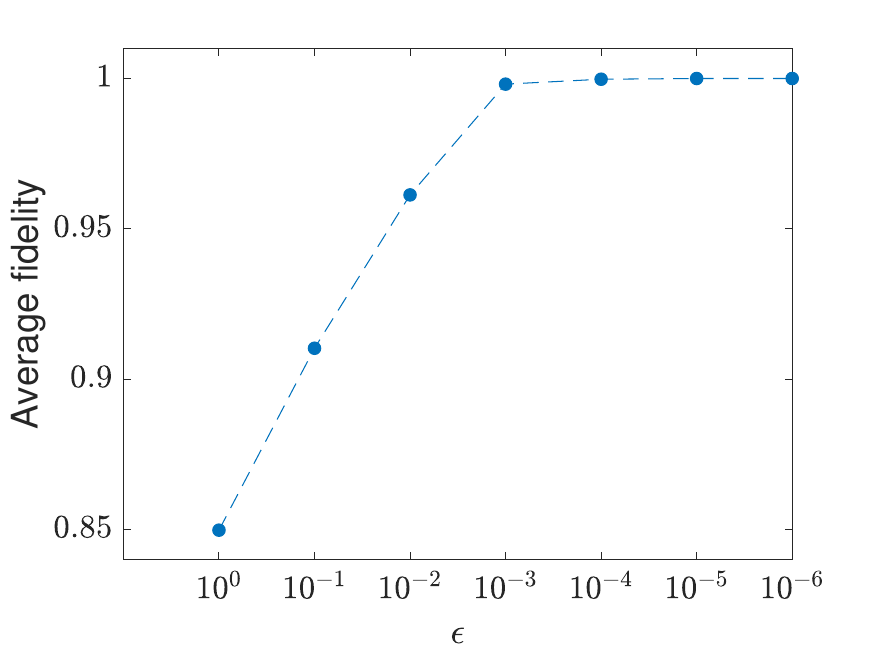}
\caption{\textbf{The impact of $\epsilon$ on the final fidelity (the 3-qubit general operator scenario).} For each $\epsilon$, we conduct 50 numerical experiments, then take the average of outcome fidelities. The vertical axis is the average fidelity and the horizontal axis is the value of $\epsilon$.}
\label{fig:epsr}
\end{minipage}
\end{figure*}

The parameter $\epsilon$, which is the halting condition of the algorithm, effects the convergence of our algorithm. It is the consequence of the object function $f(\vec{x})$ has many local minima.
In the schematic diagram~\cref{fig:mindemo}, if the initial point is chosen as $A$ or $A'$, 
the gradient method will return the value $A_0$, 
which is a local minimum (so as $B$ or $B'$ to $B_0$, and $C$ or $C'$ to $C_0$). 

When $\epsilon$ is greater then $\|f(A_0) - f(B_0)\|$ or $\|f(C_0) - f(B_0)\|$,
the algorithm may recognize $A_0$ or $C_0$ as the optimal solution instead of finding the true global minimum $B_0$. 
An appropriate choice of $\epsilon$ is necessary to eliminate certain local minima. 

The object function~\cref{eq:obj} used in our method has high dimension and complicated landscape. Its properties also subject to the particular class of $H$ that we work with. More analysis could be done in terms of finding an appropriate $\epsilon$. Empirically, we choose $\epsilon$ as $10^{-6}$ for all our numerical experiments. 
It is numerically proved to be applicable as shown in \cref{fig:epsr}.
The figure depicts the relation between $\epsilon$ and the output fidelity of the 3-qubit general operators case. Chosen $\epsilon \leq 10^{-3}$ grantees the average fidelity almost equals to 1.

By setting the halting parameter $\epsilon$ to $10^{-6}$, we can now discuss the time cost for each numerical experiment.
The total time cost $t$ for optimization depends on the number of trials $n$ for each numerical example 
and the time $t_0$ of each trial. 
The initial values are chosen randomly, $n$ is different from case to case. 
We consider the average number of trials $\bar{n}$ for each Hamiltonian class.
Let $t_0$ be the duration of a single trial which is mainly depending on the Hamiltonian configuration and number of qubits. 
The total average time $\bar{t}$ could be estimated as 
\begin{equation}
\bar{t}=\bar{n}t_0.
\end{equation} 

We test our algorithm on a workstation with Intel i7-8700K and 32 GB RAM, the results are listed in \cref{tab:performance}.
The table demonstrates that the time cost does not grow rapidly for the general operator cases, while it changes dramatically with the system size for the local operator cases.
The $\bar{t}$ of 7-qubit general operator instance is almost 2 times more then $\bar{t}$ of the 4-qubit general operator case.
On the contrast, the average time cost $\bar{t}$ of 7-qubit lattice is almost 350 times more then the $\bar{t}$ of 4-qubit lattice.

\begin{table*}[!htb]
\begin{tabular}{c|cccc|cc}
\hline
\hline
 & \multicolumn{4}{c}{General operators} & \multicolumn{2}{|c}{Local operators} \\
 \hline
Number of qubits &   4  &  5   & 6    &7    &4           &7          \\
\hline
Time for single trial $t_0$(s) &  0.0487   &    0.0518 & 0.0539    &  0.0579  &      0.2246     &     7.938     \\
 Number of trials $\bar{n}$&  32.36   &  42.34   &   45.89  &  161.6  &          54.84 &     539.4  \\
 \hline  
 Average time cost  $(\bar{t})$ (s)&  1.576   &  2.193   &   2.377 &  2.4735  &          12.32 &     4282  \\
 \hline
 \hline
\end{tabular}
\caption{\textbf{Time consumed of each trial and the number of trials.} For general operator cases and the 4-qubit local operators case, we used 200 examples to achieve the averages. For 7-qubit local operators, The result is obtained from 40 examples.}
\label{tab:performance}
\end{table*}

 \section{Comparison with the correlation matrix method}

In Ref.~\cite{qi2017determining},  a method is proposed to recover Hamiltonians from correlation functions. Their method works as follows: with an eigenstate $|\psi\rangle$, the Hamiltonian of the system is defined as 
\begin{equation}
H=\sum_i c_i L_i,
\end{equation}
where $L_i$s are normalized and orthogonal Hermitian matrices (e.g. the Pauli matrices). 
They defined a correlation matrix, of which the matrix elements are
\begin{equation}
M_{ij}=\frac{1}{2}\langle \psi |\{L_i,L_j\}|\psi\rangle-\langle \psi |L_i|\psi\rangle\langle \psi |L_j|\psi\rangle,
\label{eq:corr_mat}
\end{equation}
where $\{L_i,L_j\}=L_iL_j+L_jL_i$ is the anti-commutator. Then diagonalize matrix $M$ and find the eigenvector $\vec{\omega}=(\omega_1,\omega_2,\cdots,\omega_i)^{T}$ corresponding to the eigenvalue $0$.
The Hamiltonian
\begin{equation}
H_{cor}=\sum \omega_i L_i
\end{equation} 
is the desired Hamiltonian of the system $H$. We refer to this algorithm as the correlation matrix method (CMM).

We conduct numerical experiments to compare the performance of CMM and our method.
We calculate such 4-qubit Hamiltonians
\begin{equation}
H=\sum_{i=1}^4 a_i \sigma_z^i +\sum_{i=1}^3 b_i \sigma_x^{i}\sigma_x^{i+1},
\end{equation}
where $\sigma_k^i$ is the $k-$th Pauli matrix acting on the $i-$th qubit.
It turns out that CMM and our algorithm both render good estimations. 
The results of CMM possess greater accuracy--the error rate, defined as $r(H_\text{rd},H_\text{al}) = 1-f(H_\text{rd},H_\text{al})$, is less than $10^{-10}$.
Error rates of our results range from $10^{-4}$ to $10^{-2}$.

CMM is deterministic, more accurate as well as faster than our method.
It only involves diagonalization of matrix $M$, of which the dimension is polynomial of the number of qubits for local systems. 
CMM also provides a criteria to determine whether the eigenstate is uniquely determine the Hamiltonian. That is, if $M$ only has one eigenvalue equals to $0$, then the Hamiltonian is uniquely determined. 
The advantage of CMM arises from more restrictions and more information required for applying it:
(1) All $L_i$s in CMM are orthogonal, while in our tests the corresponding $A_i$s are randomly generated; 
(2) CMM requires non-local correlation functions. 
The matrix element includes the term
\begin{equation}
\langle \psi |\{L_i,L_j\}|\psi\rangle,
\end{equation}
where $\{L_i,L_j\}=L_iL_j+L_jL_i$. Although $L_i$s may be local, $L_iL_j$ for all $i$s and $j$s could be global. 
Therefore, as indicated in~\cite{qi2017determining}, if only partial knowledge of the system is available, the question that whether the Hamiltonian can be reconstructed remains unclear.

Another problem of applying CMM is similar to the problem of the halting parameter $\epsilon$ in our method. 
The recovery depends on the existence of one eigenvalue equals to $0$.
However, the equivalence of tow numbers in a computer is different from in theory. 
Even when working with noiseless theoretical data, the finite length data storage (e.g. the precision of double float-point format is $10^{-16}$) and data processing can introduce certain errors, not to mention the noisy data from real quantum devices.
Namely, determine whether a number is equal to $0$ is up to a certain precision. 
Therefore, one needs to set a threshold $\delta$ to determine equivalence before applying CMM.
How to appropriately chose a $\delta$ is empirical and case dependent.
From this perspective, both methods are not similarly complicated.

\begin{figure}[ht]
\includegraphics[scale=0.5]{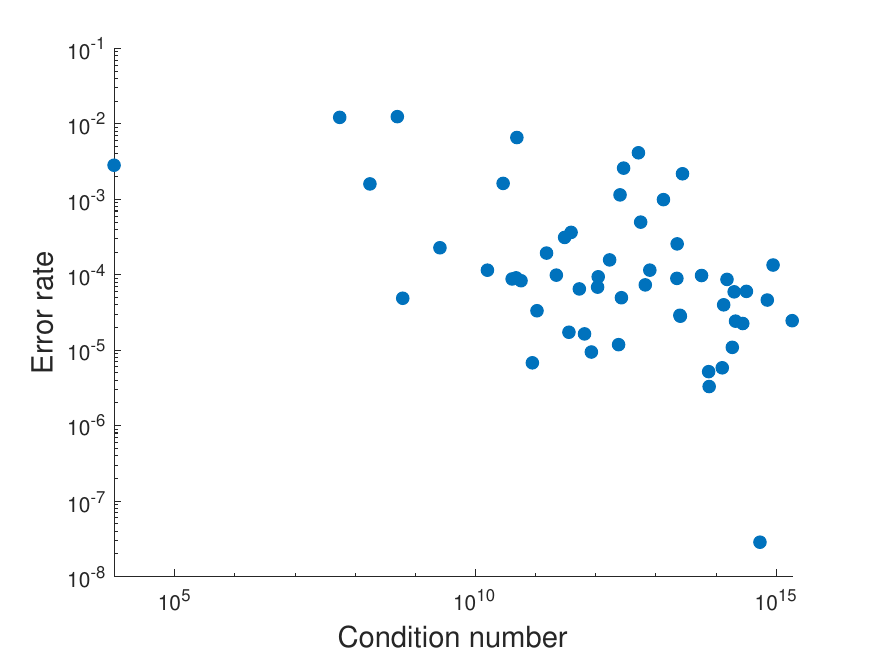}
\caption{\textbf{The impact of condition number on our algorithm.} The $x$ axis is the condition number. The $y$ axis is the error rate $r(H_\text{rd},H_\text{al}) = 1-f(H_\text{rd},H_\text{al})$.}
\label{fig:condn}
\end{figure}

We are also willing to check the consistency of CMM and our methods.
Lacking a systemic way of choosing the threshold $\delta$,
we need to seek another way to bridge these two method. 
From the numerical experiments, we find in CMM that the lowest absolute eigenvalue (denotes as $\lambda_0$) of $M$ is about $10^{-16}~10^{-18}$ and the second lowest absolute eigenvalue (denotes as $\lambda_1$) ranges from $10^{-2}$ to $10^{-12}$. If $\lambda_1$ is too small, it would be hard to tell whether there exists one or more 0 eigenvalues, that is, the farther $\lambda_0$ and $\lambda_1$ are, the more accurate  the CMM’s outcome is. 
It will be an evidence of consistency if we can witness the same tendency from our method.
We, therefore, define the \textit{condition number} as the ratio of the second lowest absolute eigenvalue to the lowest absolute eigenvalue ($\lambda_1/\lambda_0$). 
We use the estimations provided by our method to construct the correlation matrix $M$, obtain the $\lambda_0$ and $\lambda_1$ from $M$, then calculate the condition number $\lambda_1/\lambda_0$.
The results is shown in~\cref{fig:condn}.
We find, generally, the larger the condition number, the better our algorithm performs (higher fidelity). 
It is consistent with the behaviour of CMM.

In summary, for all the tests performed, CMM and our method has similar behaviour in terms of the condition number, which shows that both method return the desired results (for reconstruction the desired Hamiltonian with high fidelity).
Regarding to algorithm performance, when all sorts of correlations are available, the CMM is numerically more efficient and accurate. In contrast, however, our algorithm use less information, therefore can be used in a much wider situations, especially when certain global correlation information is hard to obtain in practice.

\section{Discussion}

In this work, we discuss the problem of reconstructing system Hamiltonian $H=\sum_i c_i A_i$ using only measurement data $a_i=\bra{\psi} A_i\ket{\psi}$ of one eigenstate $\ket{\psi}$ of $H$ by reformulating the task as an unconstrained optimization problem of some target function of $c_i$s. 
We numerically tested our method for both randomly generated $A_i$s and also the case that $A_i$s are random local operators. 
In both cases, we obtain good fidelities of the reconstructed $H_{al}$. Our results are somewhat surprising: only local measurements on one eigenstate are enough to determine the system Hamiltonian for generic cases, no further assumption is needed. Though theoretically it is beautiful, our algorithm is not scalable since the calculations of exponentiation and gradient of matrices become expensive when the system size gets large. Improvements can be done by more efficiently approximate them.

We also remark that, in the sense that our method almost perfectly recovered the Hamiltonian $H$, the information encoded in the Hamiltonian $H$, such as the eigenstate $\ket{\psi}$ itself (though described exponentially many parameters in system size) can also in principle be revealed. This builds a bridge from our study to quantum tomography and other related topics of local Hamiltonians. 
Empowered by traditional optimization methods and machine learning techniques, our algorithm could be applied in various quantum physics problems, such as quantum simulation, quantum lattice models, adiabatic quantum computation, etc.

Our discussion also raises many new questions. 
For instance, a straightforward one is whether other methods can be used for the reconstruction problem, and their efficiency and stability compared to the method we have used in this work. 
One may also wonder how the information of ``being an eigenstate" helps to determine a quantum state locally, which is generically only the case for the ground state, and how this information could be related to help quantum state tomography in a more general setting.

\section*{Aknowlegement}
We thank Feihao Zhang, Lei Wang, Shilin Huang, and Dawei Lu for helpful discussions. 
S.-Y.H is supported by National Natural Science Foundation of China under Grant No. 11847154. N.C. and B.Z. are supported by NSERC and CIFAR.


%

\appendix

\section{The value of $\beta$}
\label{sec:app}

It is easy to observe that $\tilde{H}^2$ and $\tilde{H}_{\beta}^2$ have the same eigenstates. The constant $\beta$ only contributes a constant factor to the magnitude of eigenvalues, i.e., the eigenenergies of the given system. 
Theoretically, a thermal state tends to the ground state of the given Hamiltonian if and only if the temperature is zero (or, for a numerical algorithm, close to zero), which means $\beta=\frac{1}{kT}$ goes to infinite. Numerically, the $\beta$ only needs to be a sufficiently large positive number.

\begin{figure}[h]
\subfigure[]{\includegraphics[width = \linewidth]{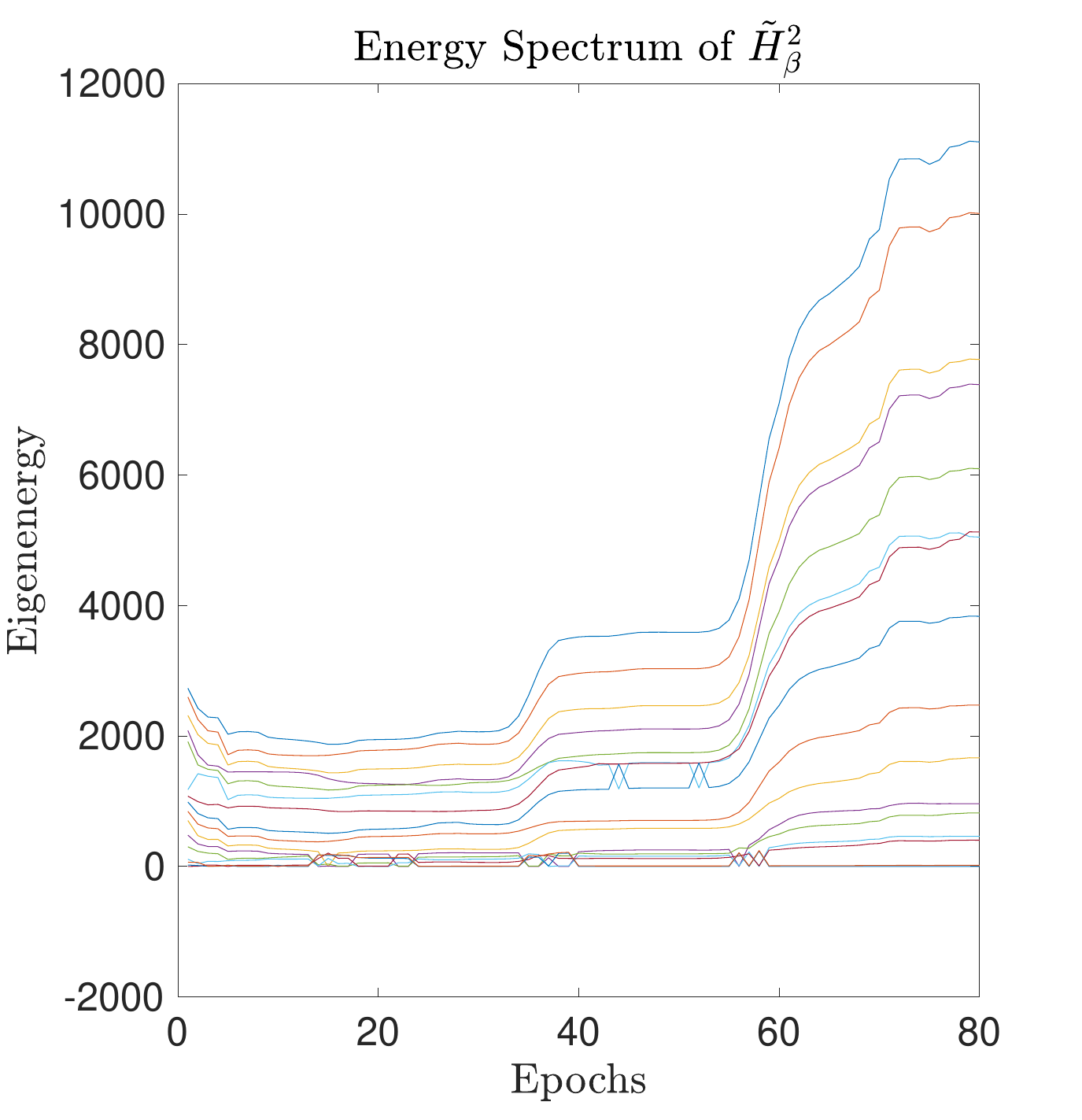}}

\subfigure[]{\includegraphics[width = \linewidth]{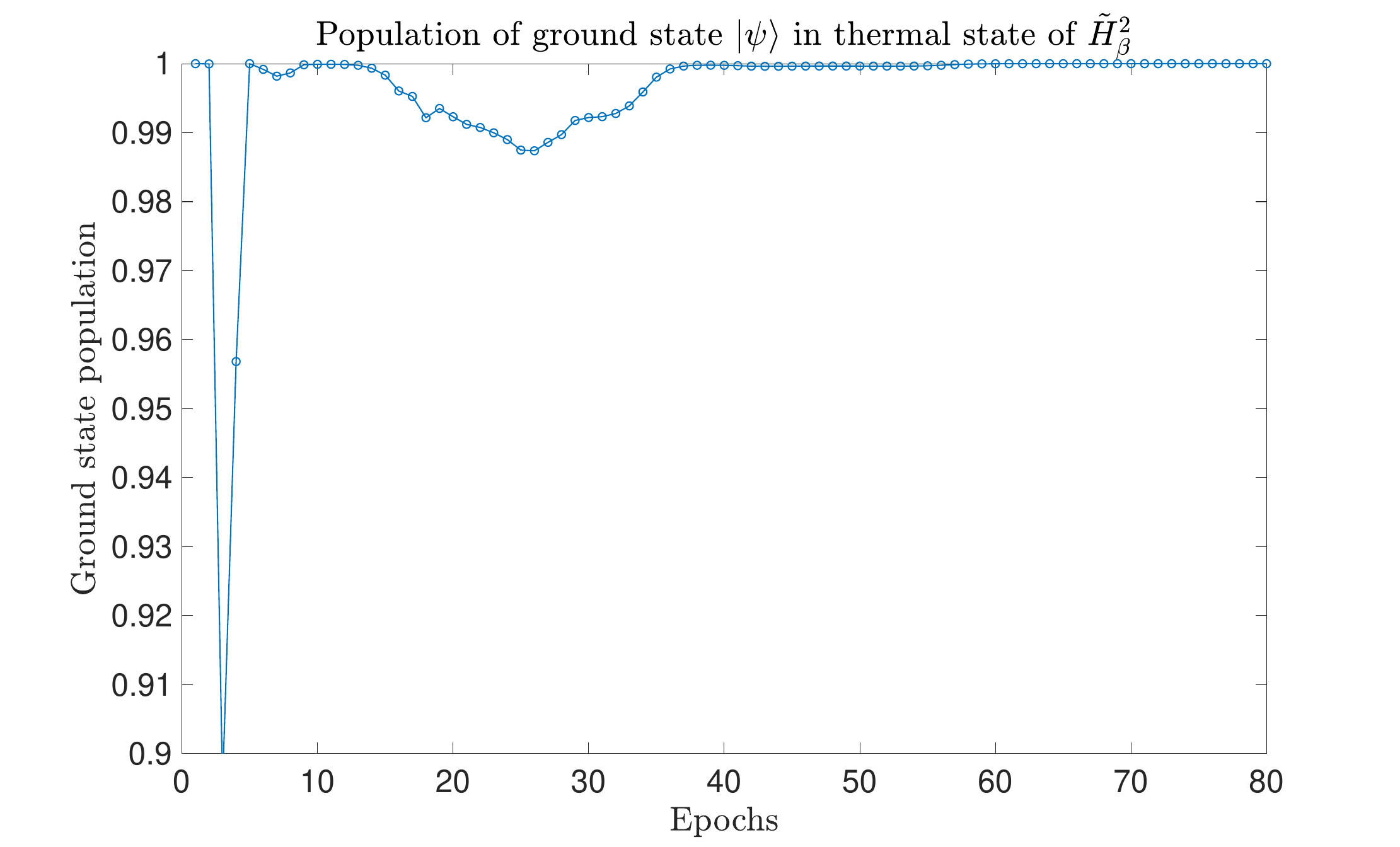}}
\caption{\textbf{Behavior of $\tilde{H}_\beta^2$ in the optimization process.} (a) Energy spectrum of $\tilde{H}^2_\beta$. (b) The ground state population during optimization.}

\label{fig:opt_proc}
\end{figure}

Let us denote the $i$-th eigenvalue of $\tilde{H}_\beta^2$ as $E_i$ and the energy gaps as $\Delta_i=E_{i+1}-E_{i}$. From the definition of $\tilde{H}_\beta^2$, the ground energy $E_g=E_1$ is always 0. As we observed, during the optimization process, the eigenenergies, as well as the energy gaps of the Hamiltonian, gets larger. 
From~\cref{fig:opt_proc}(a), we can see the energy gaps grow as the optimization goes. The corresponding $\beta$ is a finite sufficient large number.

On the other hand, we can also examine the probability of the ground state of $\tilde{H}^2_\beta$ as the iteration goes. This probability can be expressed as 
\begin{equation*}
P(E=E_g)=\frac{e^{-E_g}}{\Tr e^{-\tilde{H}_\beta^2}}
\end{equation*}
Since the ground state energy of $\tilde{H}_\beta^2$ is always have zero, the probability is 
\begin{equation*}
P(E=E_g)=\frac{1}{\Tr e^{-\tilde{H}_\beta^2}}.
\end{equation*}
The change of the probability $P(E=E_g)$ with iterations is shown in~\cref{fig:opt_proc}(b), from which we can see that finally, the probability goes to $1$, which means that the thermal state form is (almost) ground state when $\beta$ is a relatively large positive constant.

\section{Calculation of the gradient of $f(\vec{c})$}
\label{sec:grad}
\subsection{General method}

\begin{figure*}[t]
\includegraphics[scale=0.7]{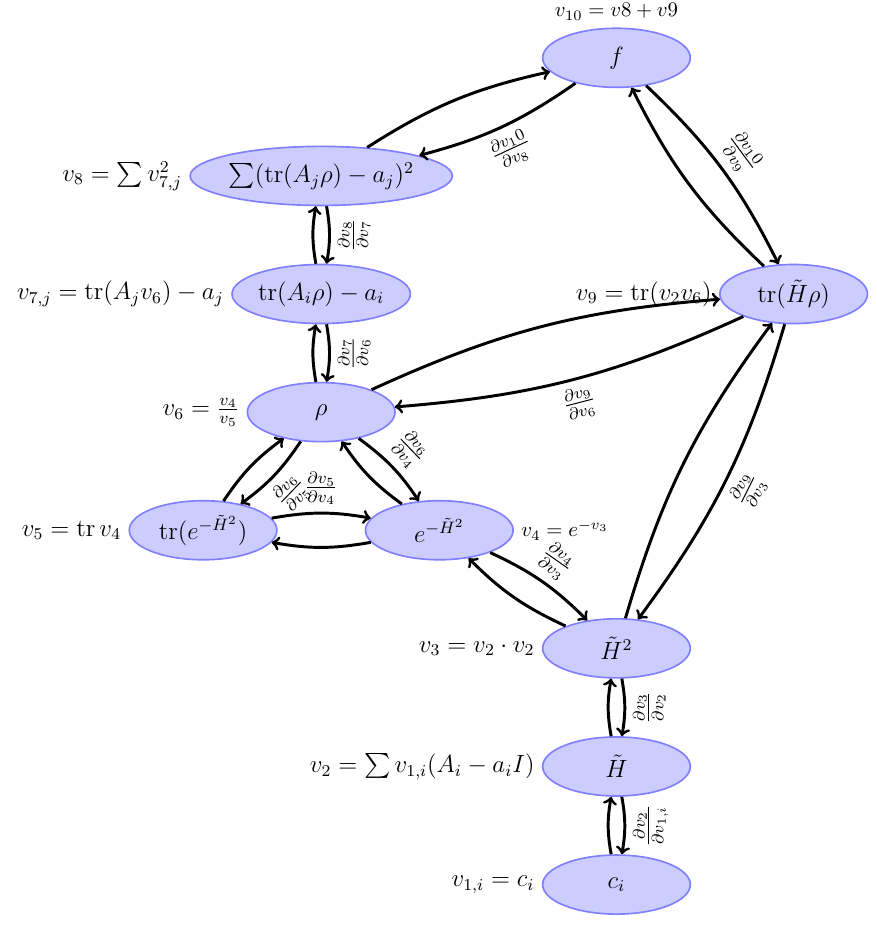}
\caption{\textbf{Computational graph.} Each node represents a intermediate function with the label on its left. The arrows from lower nodes to upper nodes represent the propagation of function while those for upper nodes to lower nodes represent the propagation of the derivatives.}
\label{fig:comgra}
\end{figure*}

The computational graph of this technique is shown in~\cref{fig:comgra}. 
Each node of the graph represents an intermediate function. 
In principle, the chain rule can be applied. It seems that we can compute the gradient by using the existing deep learning frameworks.
However, those auto differentiation tools embedded in the frameworks such as PyTorch~\cite{paszke2017automatic} and TensorFlow~\cite{tensorflow2015-whitepaper} cannot harness our problem due to the following two reasons: first, most of the tools only deal with real numbers while in quantum physics we often deal with complex numbers; 
second and the most important, some of the intermediate functions in the computational graph use matrices (or in physics, operators) as variables, while in neural networks, the variables are real numbers.
One should be careful while deriving matrices since a matrix usually does not commute with its derivative. 
For instance, even for some simple conditions, the node $v_3=\tilde{H}^2=v_2^2$, the derivative $\frac{\partial v_3}{\partial v_{1,k}}$ can not be simply considered as $2v_2\frac{\partial v_2}{\partial v_{1,k}}$ because $[\tilde{H},\frac{\partial\tilde{H}}{c_k}]\neq 0$. Instead, $\frac{\partial v_3}{\partial v_{1,k}}=v_2\frac{\partial v_2}{\partial v_{1,k}}+\frac{\partial v_2}{\partial v_{1,k}}v_2$.  

The derivatives in the computational graph shown in~\cref{fig:comgra} are listed in \cref{tab:grad}. Here, due to the reason mentioned
above, we use forward propagation to calculate the gradients. In \cref{tab:grad}, the derivative could be separated into the following categories: 1. The variable(s) of the function is a number. This corresponds to the simplest case, and the derivatives could be obtained simply using chain rules; 2. The variables are matrices, but the function is the trace function $\tr$. The derivatives could be obtained by applying $\frac{d\tr(A)}{dx}=\tr\frac{dA}{dx}$; 
3. Some elementary functions take matrices as variables, i.e., $v_3=v_2^2$. In this case, one should be careful about the commutators when dealing with it; 4. $v_4=e^{\tilde{H}^2}$, of which the gradients are difficult to calculate, which will be discussed in detail.

\begin{table*}[htb]
        \centering
       \begin{tabular}{c|c|c|c|c}

         Node & Function & Derivative & Type of variable(s) & Type of function value \\
         \hline
         \hline
         $v_2$ & $v_2=\sum c_i(A_i-a_i I)$ & 
         $A_i-a_iI$ & number & matrix
          \\
         
         $v_3$ & $v_3=v_2\cdot v_2$ & 
         $\frac{\partial v_2}{v_{1,k}}\tilde{H}+\tilde{H}\frac{\partial v_2}{v_{1,k}}$ & matrix & matrix\\
         $v_4$ & $v_4=e^{-v_3}$ & 
         $\int_0^1 e^{\alpha v_3}\frac{\partial v_3}{v_{1,k}}e^{(1-\alpha) v_3}d\alpha$ & matrix & matrix\\
         $v_5$ & $v_5=\tr v_4$ & 
         $\tr(\frac{\partial v_4}{v_{1,k}})$ & matrix & number\\
         $v_6$ & $v_6=\frac{v_4}{v_5}$ & 
         $ \frac{\partial v_4}{v_{1,k}}\frac{1}{v_5}-\frac{v_4}{v_5^2}\frac{\partial v_5}{\partial v_{1,k}}$ & matrix \& number & matrix\\
         
         $v_7$ & $v_7,j=\tr(A_j v_6) - a_j$ & 
         $\tr(A_j \frac{\partial v_6}{\partial v_{1,k}}) $ & matrix & number\\
         
         $v_8$ & $v_8=\sum v_{7,j}^2$ & 
         $2v_{7,j}\frac{\partial v_{7,j}}{\partial v_{1,i}} $ & number & number\\
         $v_9$ & $v_9=\tr(v_3 v_6)$ & 
         $\tr(\frac{\partial v3}{\partial v_{1,i}}v_6 +v_3\frac{\partial v6}{\partial v_{1,i}}) $ & matrix & number\\
         $v_{10}$ & $v_10=v_8+v_9$ & $\frac{\partial v8}{\partial v_{1,i}}+\frac{\partial v9}{\partial v_{1,i}}$ & number & number\\
         \hline
         
       \end{tabular}
       
       \caption{\textbf{The types of the variables and values of the intermediate functions.} }\label{tab:grad}
\end{table*}

\subsection{Derivative of matrix exponentials}

The derivative of function $f(X)=\exp(X(c_k))$ with respect to $c_k$ can be written as \begin{equation}
\frac{\partial e^{X(c_k)}}{\partial c_k}=\int_0^1 e^{\alpha X(c_k)}\frac{\partial X(c_k)}{\partial c_k}e^{(1-\alpha)X(c_k)}d\alpha,
\label{eq:gradexp}
\end{equation}
where, in our cases, $X(c_k)=-\tilde{H}(c_k)^2=-v_3$. Generally, the commutator  $[X,\frac{\partial X}{\partial c_k}]\neq 0$. 
Therefore, the integration \cref{eq:gradexp} can not be simply calculated. 
In Ref.~\cite{niekamp2013computing}, the authors approximate the integration using the value of the upper and lower limits.
This approach does not work for our case. 
Thus we introduce a new way to calculate it.
Let\begin{equation}
A(\beta)=\frac{\partial e^{X(c_k)}}{\partial c_k}=\int_0^\beta e^{\alpha X(c_k)}\frac{\partial X(c_k)}{\partial c_k}e^{-\alpha X(c_k)}d\alpha,
\end{equation}

\begin{equation}
B(\beta)=e^{-\beta X(c_k)}A(\beta),
\end{equation}
and
\begin{equation}
C(\beta)=e^{-\beta X(c_k)}.
\end{equation}
Note that $A(1)=B(1)\exp[X(c_k)]$.
It can be derived that 
\begin{equation}
\frac{d}{d\beta}\left(\begin{matrix}
B(\beta)\\
C(\beta)
\end{matrix}\right)
=\left(\begin{matrix}
-X(c_k) & \frac{\partial X(c_k)}{\partial c_k} \\
0 & -X(c_k)
\end{matrix}\right)
\left(\begin{matrix}
B(\beta)\\
C(\beta)
\end{matrix}\right).
\label{eq:deriv}
\end{equation}

Let
\begin{equation}
G=\left(\begin{matrix}
-X(c_k) & \frac{\partial X(c_k)}{\partial c_k} \\
0 & -X(c_k)
\end{matrix}\right),
\end{equation}
The derivative equation \eqref{eq:deriv} can be solved as
\begin{equation}
\left(\begin{matrix}
B(1)\\
C(1)
\end{matrix}\right)
=e^G
\left(\begin{matrix}
B(0)\\
C(0)
\end{matrix}\right),
\end{equation}
where $B(0)$ is a matrix with all entries 0 and $C(0)$ is the identity matrix. With $B(1)$, we can obtain $A(1)$ as well as the integration~\cref{eq:gradexp}.
This completes the gradient calculation for our algorithm.

\section{Software implementations}

According to the computational graph \cref{fig:comgra} and the \cref{tab:grad}, we can define a function which accepts $\vec{c}$, the operators $A$, and returns the value of $f(\vec{c})$ and the gradient $\nabla f(\vec{c})$:
\begin{lstlisting}[style=matlab-editor]
function [f,g]=fun(c, A)
% f is the value of the objective function
% g is the gradient
\end{lstlisting}

 The function \textsc{fminunc} in MATLAB/Octave can accept the function and the gradient and doing the optimization:
\begin{lstlisting}[style=matlab-editor]
f_handle=@(c)fun(c,A)
\end{lstlisting}
 The default algorithm is BFGS. To make the algorithm using gradient we provide:
\begin{lstlisting}[style=matlab-editor]
options=optimoptions('fminunc','SpecifyObjectiveGradient','true');
\end{lstlisting}
Then, we can start the optimization
\begin{lstlisting}[style=matlab-editor]
[c1, fval]=fminunc(f_handle,c0)
\end{lstlisting}
Here, c1 is the final points, fval is the final value of the objective function and c0 is the initial value.

\end{document}